\documentclass[aps,amssymb,amsmath,preprint, superscriptaddress, pra]{revtex4}
\usepackage{graphicx}
\usepackage{epsfig}
\usepackage{dcolumn}
\usepackage{bm}
\usepackage{amsfonts}
\usepackage{color}
\usepackage{hyperref}
\usepackage{epstopdf}
\usepackage[vcentermath]{youngtab}
\usepackage{young}

\newcommand{\be}{\begin{equation}}
\newcommand{\ee}{\end{equation}}
\newcommand{\bc}{\begin{center}}
\newcommand{\ec}{\end{center}}
\newcommand{\bea}{\begin{eqnarray}}
\newcommand{\eea}{\end{eqnarray}}

\newcommand{\ket}[1]{\left|#1\right.\rangle}

\newcommand{\TK}[1]{{ #1}}

\definecolor{orange}{rgb}{1,0.5,0}

\begin{document}
\title{Teleporting photonic qudits using multimode quantum scissors}

\author{Sandeep K. \surname{Goyal}}
\affiliation{School of Chemistry and Physics, University of KwaZulu-Natal, Private Bag X54001, Durban 4000, South Africa}

\author{Thomas \surname{Konrad}}
\email{konradt@ukzn.ac.za}
\affiliation{School of Chemistry and Physics, University of KwaZulu-Natal, Private Bag X54001, Durban 4000, South Africa}  
\affiliation{National Institute of Theoretical Physics (NITheP),  University of KwaZulu-Natal, Private Bag X54001, Durban 4000, South Africa}

\begin{abstract} 
 Teleportation plays an important role in the communication of quantum information between the nodes of a quantum network and is viewed as an essential ingredient for long-distance Quantum Cryptography. We describe a method to  teleport  the quantum information  carried by a photon in a superposition of a number $d$ of  light modes  (a  ``qudit'') by the help of  $d$ additional  photons based on transcription.  A qudit encoded into a single excitation of $d$ light modes (in our case Laguerre-Gauss modes which carry orbital angular momentum)  is transcribed to $d$ single-rail photonic qubits, which are spatially separated.  Each single-rail qubit consists of a superposition of vacuum and a single photon in each one of the modes. After successful  teleportation of each of the  $d$ single-rail qubits by means of ``quantum scissors''  they are converted back  into a qudit carried by a single photon which completes the teleportation scheme.
 \end{abstract}

\maketitle

  Quantum teleportation, the carrier-less transmission of quantum information by transferring a state from one  quantum system to a remote one  was described by Bennett {\em et al.} \cite{Bennett1993}  and soon after played an important role in photonic quantum computing \cite{Brassard1998, Gottesman1999, Knill2001, Duan2001} as well as secure communication by means of quantum key distribution (e.g.\  \cite{Briegel1998,Tittel2000}). 
The fragile nature of quantum systems and nearly omnipresent dissipative environments make it challenging to realize quantum teleportation experimentally. Bouwmeester {\em et al.} \cite{Bouwmeester1997} were the first to achieve quantum teleportation followed by many others {for discrete-level quantum systems \cite{Marcikic2003,Boschi1998,Nielsen1998,Jennewein2001,Kim2001,Lombardi2002,Babichev2003} as well as with continuous variables  \cite{Furusawa1998,Bowen2003,Takei2005,Sherson2006}. In the case of discrete-level quantum systems} so far only the state of the most simple  quantum systems, i.e.,  two-level systems,  and therewith the smallest unit of quantum information (a ``qubit'') could be teleported.  A new teleportation scheme, proposed recently \cite{Goyal2012}, is capable to transmit  the quantum information carried by an elementary excitation of a superposition of an arbitrary number $d$ of  co-propagating light modes  (a {photonic} ``qudit''). {The teleportation of photonic qudits increases the quantum information sent per carrier photon. Currently, the low transmission rates are one of the bottlenecks of quantum communication as compared to its classical counterpart.}  However, the scheme proposed in \cite{Goyal2012} requires to prepare $d$ additional photons in a highly entangled state.  Here we present an alternative scheme  based on the transcription of the qudit encoded in a single photon  to $d$ qubits carried by light modes which propagate along different optical paths.  Each qubit contains the quantum information about the excitation of a particular of the  $d$ original light modes and is  teleported individually by means of an additional photon  using quantum scissors.

\begin{figure}
\includegraphics[width=8.5cm]{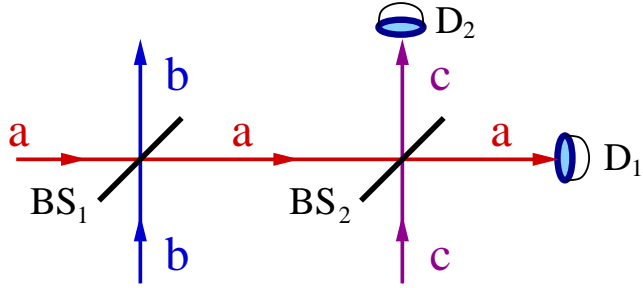}
\caption{Schematic diagram for  quantum scissors. In the quantum scissors setup there are two $50:50$ beam splitters (BS$_1$ and BS$_2$)  and two number-resolving photon detectors ($D_1$ and $D_2$). An optical state $\ket{\chi}$ is inserted in mode $c$ along with a single photon in mode $a$ and vacuum in $b$. Beam splitter BS$_1$ entangles mode $a$ and $b$ by distributing the incoming photon. 
A detection of a single photon in  $D_1$ or $D_2$ leaves mode $b$ in a superposition of vacuum (in case the detection annihilated  the photon in mode $a$) and a single photon state (in case the detected light originated not from mode $a$ but $c$). The superposition state in $b$ is caused by beam splitter BS$_2$, which deletes the path information about the origin of the detected light.}\label{q-scissors}
\end{figure}

Quantum scissors \cite{Pegg1998, Barnett1999, Ozdemir2001, Ozdemir2002, Babichev2003}
is a device to teleport only the vacuum and the single-photon component of a single-mode state (a so-called single-rail qubit), while it truncates (``cuts off'')   higher photon-number components. In particular, if an input light mode $c$ (cp. Fig.~\eqref{q-scissors}) is prepared in a superposition of states \TK{ with different photon numbers},  quantum scissors projects its vacuum and single photon component to an ouput mode $b$:
\begin{equation}
\ket{\chi}_c=( \alpha_0 \ket{0}_c + \alpha_1\ket{1}_c + \alpha_2\ket{2}_c +\ldots)\,\, \to \ket{\chi'}_b=\frac{1}{2}\left( \alpha_0\ket{0}_b + \alpha_1 \ket{1}_b \right)\,,
\label{initial}
\end{equation}
where $\ket{n}_{c (b)}$ represents a so-called {\sl Fock state} with $n=0,1,2\ldots$ photons in light mode $c$ ($b$)  and  the coefficients $\alpha_n $ are the corresponding probability amplitudes.  
This  process occurs upon conditioning on a single-photon detection with the probability given by  the square of the norm of the final state $\ket{\chi'}_b$, i.e, $(|\alpha_0|^2 + |\alpha_1|^2)/4$. {However, this probability can be doubled by conditioning on one of two possible single-photon detections (cp.\ Methods). }   The working principle of quantum scissors is explained in the caption of Fig.~\eqref{q-scissors}.

On the other hand, if the input state in mode $c$  consists already of a single-rail qubit, i.e.\ $ \ket{\chi}_c = \left( \alpha_0\ket{0}_c+\alpha_1 \ket{1}_c\right)$, it is transferred according to transformation \eqref{initial} without truncation, and hence teleported, to mode $b$. {A generalization of quantum scissors  which cuts off all state components with a number of $d$ or more photons and thus teleports multi-photon states  of the form $ \ket{\chi}_c = \sum_{n=0}^{d-1} \alpha_n\ket{n}_c$ ( ``single-rail qudits'')  can be achieved using multiports and $d-1$ additional input photons \cite{Miranowicz2005, Miranowicz2007}. However, the encoding of an arbitrary superposition $ \ket{\chi}_c$ of multiple photon-number Fock states is in practice difficult and requires non-linear optical media leading to small efficiencies \cite{Leonski1994, Leonski1997, Miranowicz1996}}.

Moreover, one can teleport $n$ single-rail qubits simultaneously, provided they are stored in light modes which propagate on different paths,  by applying $n$ quantum scissor setups in parallel, one for each single-rail qubit. Obviously, the simultaneous teleportation works if the  single-rail qubits in the individual modes are not correlated. But note, that also the state of $n$ entangled qubits can be teleported in this way. 

\section*{Results}
This feature of quantum scissors can be exploited to teleport a qudit  encoded into a single photon which is shared by $d$ spatial modes of paraxial light. For this purpose the $d$-level state of the photon is {\sl transcribed} into $d$ single-rail qubits carried by light modes propagating along different paths with the help of a mode sorter. Such a device has the task to transfer orthogonal light modes within a single light beam to different optical paths, similar to a polarizing beam splitter, which conveys light with horizontal and  vertical polarization to orthogonal paths.  For example, consider a single-photon state $\ket{\chi}$ given by  an elementary excitation of a superposition of $d$ paraxial Laguerre-Gauss modes $LG_{l,p=0}$ corresponding to different values $l\hbar$ of orbital angular momentum (OAM) \cite{Allen1992} which co-propagate   along an optical path $o$, i.e,
\begin{align}
\ket{\chi}_o &= \sum_{l=0}^{d-1} \gamma_l \ket{1_l}_o\quad\mbox{with}\quad \sum_l|\gamma_l|^2 = 1\,,
\label{eta}
\end{align}
where $\ket{1_l}$ denotes the state of a single photon with OAM  $l\hbar$. We spatially separate the OAM modes  by diverting them into different optical  paths $c_l$ depending on their OAM value $l\hbar$ with the help of an OAM mode sorter \cite{Leach2002,Berkhout2010}. For $d=3$ this transformation reads:
\begin{align}
\ket{\chi}_o &\to\ket{\chi}_{c_0\,c_1\,c_2} = \gamma_0 \ket{1_{c_0}\,0_{c_1}\,0_{c_2}} + \gamma_1 \ket{0_{c_0}\,1_{c_1}\,0_{c_2}}+ \gamma_2 \ket{0_{c_0}\,0_{c_1}\,1_{c_2}} \equiv\sum_{l=0}^{2} \gamma_l c_l^\dagger\ket{0},
\label{eta2}
\end{align}
\noindent
where $\ket{1_{c_0}\,0_{c_1}\,0_{c_2}} $ represents a single photon with OAM \TK{quantum number} $l=0$ in path $c_0$ and no photon in all other paths (accordingly for the remaining terms). The single photon states are conveniently expressed by the creation operators $c^\dagger_l$ acting on the global vacuum state $\ket{0}$, cp.\ the right-hand side of \eqref{eta2}. This transformation {\sl transcribes} the state of a qudit into $d$ entangled single-rail qubits. 
After the transcription the $i$th qubit contains the quantum information about whether the corresponding OAM mode $l=i$ of the photonic qudit was occupied ($\ket{1_i}_o\rightarrow\ket{1_{c_i}}$) or not ($\ket{0_i}_o\rightarrow\ket{0_{c_i}}$). 

\begin{figure}
\includegraphics[width=8.5cm]{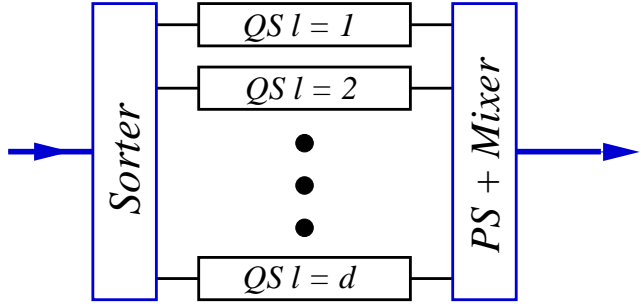}
\caption{Qudit teleportation setup  using multiple quantum scissors. The $d$-dimensional state of a single photon, which is in a superposition of $d$  OAM modes, is mapped by a mode sorter to $d$ single-rail qubits. This is followed by the teleportation of the individual single-rail qubits using $d$ quantum scissors. Eventually, a $\pi$ phase shift ($PS$) is applied to the single-rail qubits where necessary and their composite state is transcribed back into a photonic qubit by a mixer.}\label{tele}
\end{figure}

\begin{figure}
\includegraphics[width=8cm]{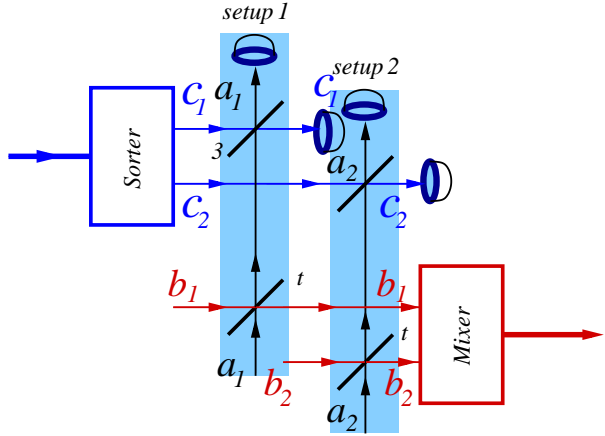}
\caption{Schematic diagram of teleportation of a photonic qubit. In this setup, the {\em mode sorter} transfers the input qubit encoded in two basis modes of the input light beam on the left to different paths $c_1$ and $c_2$. Two quantum scissors devices  teleport the states of light of the spatially separated modes $c_1$ and $c_2$ individually to $b_1$ and $b_2$, respectively. The latter are retransferred  by a mixer into a photonic qubit carried by a single output beam on the right, which is a sorter run in reverse.}\label{qubit-qs}
\end{figure}

Now each single-rail qubit can be teleported individually (cp. Fig.~\eqref{tele})
 using quantum scissors. This is accomplished as follows: each of the $d$ spatial modes are inserted into $d$ independent quantum scissors setups (see Fig.~\eqref{qubit-qs} for $d=2$). There are thus   $d$  input modes $c_i$ and  $d$ output modes $b_i$  with $i=0,1\ldots d-1$ to carry the quantum information. In addition, the quantum scissors require a total of $d$ single photons entered separately in modes $a_i$. 
Upon conditioning on the detection of a single photon in each of the quantum scissor devices (success probability $1/2^d$ {with ideal detectors, for non-unit detection efficiencies see Methods}) the state $\ket{\chi}$ carried by the input modes $c_i$ is transferred to the output modes $b_i$ (cp.\ Methods):
\begin{equation}
\ket{\chi}_{c_0\dots c_{d-1}}\rightarrow \ket{\chi}_{b_0\dots b_{d-1}}\,.\label{transfer}
\end{equation}

Since the mode $b_i$ in the $i$th quantum scissors device originates from the reflection of  mode $a_i$ both are identical except for their propagation direction, and they should carry the same OAM value as input mode $c_i$, the state of which is supposed to be transferred to $b_i$. Hence, the photon entering in mode $a_i$  should be prepared with OAM value $i \hbar$. 

 However, as shown under Methods, this is not necessary if the modes $b_i$ are transformed into the appropriate OAM mode after the state transfer. In fact, preparing the additional photons in a different system of basis modes enables a transcription of the quantum information stored in a specific basis (here OAM modes) in the input modes of the quantum scissors devices to another basis (for example Hermite Gaussian modes \cite{Leary2009}) in its output modes. {By such a transcription any unitary gate acting on the Hilbert space of the qudit  can be realized, however only with  limited success probability which is  determined by the quantum scissors involved.}

{After successful teleportation by the quantum scissors} we can convert the {entangled} $d$ single-rail qubits back to the original $d$-mode OAM state \eqref{eta} with the help of a mixer which is a sorter run in reverse.  This completes the teleportation of a photonic qudit (cp. Fig.~\eqref{tele}). { In order to realize an additional  unitary qudit-gate  (see above) together with  the teleportation,  the mixer has to map the new basis modes in the outputs of the $d$ quantum scissors into a single beam, i.e., it must be a reverse sorter for these particular modes, which exists for example for Hermite Gaussian modes \cite{Leary2009}. }        

\section*{Discussion}
In this article we have presented a scheme to teleport a photonic qudit carried by OAM modes.  The scheme requires linear optical devices, OAM mode sorters as well as single-photon sources and photon-number resolving detectors. The essential step is to transcribe the state of the qudit to $d$ single-rail qubits by means of a mode sorter and to teleport the qubits individually by quantum scissors. In as far as such sorter devices can be designed for other light modes, for example Hermite-Gaussian modes \cite{Leary2009}, the proposed teleportation scheme  is universal and can be implemented with any system of basis modes. Using quantum scissors a single-rail qubit can be teleported with success probability $1/2$, therefore, {the success probability to teleport $d$ single-rail qubits and thus the encoded qudit amounts to $1/2^d$.}

{In principle, the probability to teleport a single-rail qubit  can be increased to $N/(N+1)$ by employing $N$ additional entangled photons and a balanced multiport with $N+1$ inputs and outputs as described by Knill et al.\cite{Knill2001} instead of one additional photon and a balanced beam splitter in each of the quantum scissors setups. This  leads to a success probability for the qudit teleportation of $(N/(N+1))^d$ but requires $d$ highly entangled $N$-photon states, which can be prepared probabilistically offline\cite{Knill2001, Kok}.}

{On the other hand, the encoding of one qudit into $d$ two-level systems (such as single-rail qubits)  represents an inefficient use of storage capacity. The amount of quantum information present in a single qudit actually corresponds to $\log_2d$ qubit units of quantum information and could thus be stored efficiently in $\log_2d$ single-rail qubits. Given a scheme which is able to transcribe the initial photonic qudit into $\log_2d$ single-rail qubits, a subsequent teleportation could be achieved by means of $\log_2d$ quantum scissors with a success probability of $1/2^{\log_2d} = 1/d$. This would mean an exponential decrease of the resources needed to teleport a qudit. }

Alternatively, using the transcription based on a {OAM mode sorter, as described above,  the present scheme allows, instead of a single qudit,} to teleport $d$ single-rail qubits encoded in co-propagating {OAM modes, with the same resources as before.  This corresponds to an exponential increase of quantum information sent per use of the teleportation protocol. } However, the {preparation and} manipulation of single-rail qubits seems problematic 
 compared to qudits carried by  single-photon states of OAM modes, which can be prepared, transformed and measured with standard techniques \cite{Terriza2001}. 
{For single-rail qubits, general deterministic single- and two-qubit gates are not available \cite{Wu2013}. Moreover, the vacuum component makes state tomography of single-rail qubits difficult.}

The present scheme has certain advantages as well as disadvantages over a recently proposed alternative teleportation method \cite{Goyal2012}.  Unlike the latter, it {\em does not} require highly sensitive multipartite entangled states to perform the quantum teleportation. {On the other hand, } the alternative method yields a greater success probability of $1/d^2$ for qudit teleportation, and requires less additional photons. {Remarkably, it yields for the joint teleportation of the state of many photons the same maximal teleportation rate as quantum scissors for single rail qubits, namely one qubit per additional photon.} However, by improving the transcription efficiency one could overcome these drawbacks of the present scheme.

\section*{Methods}
In the following we show  that $d$ quantum scissors enable a transfer of the state $\ket{\chi}_{c_0\dots c_{d-1}}$ obtained after the transformation  \eqref{eta2} to output modes $b_0\ldots b_{d-1}$ \eqref{transfer}.  The state $\ket{\chi}_{c_0\dots c_{d-1}}$  along with  $d$  photons at ports $a_i$  constitute the initial state entering the $d$ quantum scissors (cp.\ Fig.~\eqref{qubit-qs} for $d=2$)
\begin{align}
\ket{1_{a_0}\ldots1_{a_{d-1}}}\otimes \ket{\chi}_{c_0\dots c_{d-1}} = \prod_{i=0}^{d-1} a_i^\dagger \sum_{l=0}^{d-1}\gamma_l c_l^\dagger \ket{0}\,,
\label{eq-9}
\end{align}
where we have introduced the creation operators $a^\dagger_i$ to denote  single photons in the modes $a_i$. 
Each quantum scissors \TK{device} contains two $50:50$ beam splitters, cp.\ Fig.\ \ref{q-scissors}.  The first beam splitter BS$_1$ of the $i$th \TK{device }distributes the incoming photon in mode $a_i$ equally over both modes, $a_i$ and $b_i$, represented by the transformation rule in terms  of the corresponding  creation operators 
$a^\dagger_i \to\frac{1}{\sqrt{2}}\left(a^\dagger_i + b^\dagger_i\right)$.
 Also the action of the second beam splitter BS$_2$ in each quantum scissors device is conveniently described by similar rules:
 \begin{align}
&a^\dagger_i \to \frac{1}{\sqrt{2}}\left(a^\dagger_i + c^\dagger_i\right),\\
&c^\dagger_i \to \frac{1}{\sqrt{2}}\left(a^\dagger_i - c^\dagger_i\right)\,. \label{second}
\end{align}
Consecutive application of these transformations for beam splitters BS$_1$ and BS$_2$ for all quantum scissors $i$ to the initial state \eqref{eq-9} yields the total state change: 
\begin{align}
\ket{1_{a_0}\ldots1_{a_{d-1}}}\otimes \ket{\chi}_{c_0\dots c_{d-1}} &\to \ket{\Phi}=  \prod_{i=0}^{d-1} \frac{1}{\sqrt{2}}\left( \frac{1}{\sqrt{2}}\left(a^\dagger_i + c^\dagger_i\right) + b^\dagger_i\right) \sum_{l=0}^{d-1} \gamma_l\left( \frac{1}{\sqrt{2}}(a^\dagger_l - c^\dagger_l)\right)\ket{0}.\label{eq-11a}
\end{align}
The second and final step to complete the state transfer by means of quantum scissors provides a photon-number measurement in modes $a_i$ and $c_i$ conditioned on the detection of a single photon in $a_i$ and vacuum in $c_i$, cp.\ Fig.\ \ref{q-scissors}. 
Since there are a total of $d+1$ photons in the system, a detection of one photon in each of the $d$ modes  $a_i$ and zero photons in the modes $c_i$ results in a single photon in one of the modes $b_i$ according to photon-number conservation. 
 The  measurement  projects onto those   components of state $\ket{\Phi}$  in 
 {Eq.~(\ref{eq-11a})}   which allow for such a detection event:
 \begin{align}
\ket{\Phi}\to& \left(\gamma_1a_2^\dagger\cdots a_d^\dagger b_1^\dagger a_1^\dagger  + \gamma_2 a_1^\dagger \cdots a_d^\dagger  b_2^\dagger a_2^\dagger \cdots + \gamma_d a_1^\dagger \cdots a_{d-1}^\dagger  b_d^\dagger a_d^\dagger\right) \ket{0} \nonumber \\
&=   a_1^\dagger a_2^\dagger \cdots a_d^\dagger \sum_l \gamma_l b_l^\dagger \ket{0} =  \ket{1_{a_0}\ldots1_{a_{d-1}}}\otimes \ket{\chi}_{b_0\dots b_{d-1}}\,, \label{eq-11}
\end{align}
Therefore the state of light in the output modes $b_0 \ldots b_{d-1}$ of the quantum scissors reads:
\begin{align}
\ket{\chi}_{b_0 \ldots b_{d-1}}&= \sum_l\gamma_l b_l^\dagger \ket{0}
\end{align}
which is the state initially carried by the input modes $c_l$, cf.~\eqref{eq-9}.

{Please note, that a teleportation of the initial state carried by the input modes $c_i$ onto different output modes $\tilde b_i$ can also be achieved:
\begin{align}
\ket{\chi}_{c_0 \ldots c_{d-1}} \rightarrow  \ket{\chi}_{{\tilde b_0} \ldots {\tilde b_{d-1}}}&= \sum_l\gamma_l \tilde b_l^\dagger \ket{0}
\end{align}
 For this purpose, photons of modes $\tilde a_i$ corresponding to the targeted modes $\tilde b_i$ are inserted into the ports $a_i$, together with the initial state $\ket{\chi}$ in modes $c_i$; 
 \begin{align}
\ket{1_{\tilde a_0}\ldots1_{\tilde a_{d-1}}}\otimes \ket{\chi}_{c_0\dots c_{d-1}} = \prod_{i=0}^{d-1} \tilde a_i^\dagger \sum_{l=0}^{d-1}\gamma_l c_l^\dagger \ket{0}\,.
\label{eq-9b}
\end{align}
The consecutive actions of the  beam splitters $BS_1$ and $BS_2$ in the quantum scissor devices, given respectively by $\tilde a^\dagger_i \to\frac{1}{\sqrt{2}}\left(\tilde a^\dagger_i + \tilde b^\dagger_i\right)$ and $\tilde a^\dagger_i \to \frac{1}{\sqrt{2}}\left(\tilde a^\dagger_i + \tilde c^\dagger_i\right)$ together with \eqref{second}, transform state \eqref{eq-9b} into 
\begin{align}
 \ket{\Psi}=  \prod_{i=0}^{d-1} \frac{1}{\sqrt{2}}\left( \frac{1}{\sqrt{2}}\left(\tilde a^\dagger_i + \tilde c^\dagger_i\right) + \tilde b^\dagger_i\right) \sum_{l=0}^{d-1} \gamma_l\left( \frac{1}{\sqrt{2}}(a^\dagger_l - c^\dagger_l)\right)\ket{0}\,.\label{eq-11b}
\end{align}
The only components of state $\ket{\Psi}$ that can contribute to a coincidence detection of a single photon by the detectors  $D_1$   (cp. Fig.~\ref{q-scissors}) in each of the $d$ quantum scissors devices are given by  
 \begin{align}
\left(\gamma_1\tilde a_2^\dagger\cdots \tilde a_d^\dagger \tilde b_1^\dagger a_1^\dagger  + \gamma_2 \tilde a_1^\dagger \cdots \tilde a_d^\dagger  \tilde b_2^\dagger a_2^\dagger \cdots + \gamma_d \tilde a_1^\dagger \cdots \tilde a_{d-1}^\dagger  \tilde b_d^\dagger a_d^\dagger\right) \ket{0} 
\end{align}
If we further assume that each detectors $D_1$ absorbs a single photon in the detection process without distinguishing between both kinds of photons \cite{Ozdemir2002}, $a_i$ and $\tilde a_i$, then the remaining state is given as claimed by
\begin{align}
 \ket{\chi}_{{\tilde b_0} \ldots {\tilde b_{d-1}}}&= \sum_l\gamma_l \tilde b_l^\dagger \ket{0}\,.
\end{align}
}

The {detection event indicating successful teleportation occurs for ideal detectors} with probability $1/2^{2d}$ which is obtained from a normalization factor in projection~\eqref{eq-11}. However, the success probability can be increased by considering other detection events. 
For example, if  one photon is detected in mode $c_j$ instead of mode  $a_j$, as well as one photon in each of the remaining modes $a_i$,  the state of  $b$ collapses into:{
\begin{align}
\ket{\bar{\chi}}_b &= \sum_l\bar{\gamma}_l b_l^\dagger \ket{0},
\end{align}}
where {$\bar{\gamma}_l$ is $\gamma_l$} for $l \ne j$  and $-\gamma_l$ for $l = j$. The minus sign can be compensated by applying a $\pi$-phase shift to mode $b_j$ which causes the state change {$\ket{\bar{\chi}}_b \to \ket{\chi}_b$.} Hence, it does not matter whether the detectors in modes $a_j$  or $c_j$ register a single photon count as long as there is only one count  in each quantum scissors setup. Thus there are  $2^d$ detection events corresponding to successful teleportation, which increases the probability of success to $2^d/2^{2d}=1/2^d$. 

{The success probability of the teleportation scheme depends on the efficiencies of the detectors used with the quantum scissors. For detectors which count a single photon with probability (i.e., efficiency) $\eta <1$ the success probability of the scheme reduces to $(\eta/2)^d$. Moreover, a restricted detection efficiency can induce the false identification of a two-photon detection event as a single-photon count with probability $2\eta(1-\eta)$. Such a mistaken identification in one of the quantum scissor set-ups together with single photon counts in the remaining ones leads to vacuum in the output modes, while the detectors seemingly announce a successful teleportation. The probability for such false announcement equals $\eta(1-\eta)$ and can be calculated from the  probability to obtain a two-photon detection event with ideal detectors which amounts to $1/2$, independent of the number of quantum scissor setups (cp.\ Eq.\ \eqref{eq-11a}),  multiplied with the probability for a false identificaton due to the non-unit detection efficiency.    Therefore, the teleportation fidelity defined as the overlap between the input and the output state of the teleportation scheme decreases from $f = 1$ with ideal detectors to  $f = 1-\eta(1-\eta)$ for detectors with efficiency $\eta$.}

\makeatletter
\renewcommand\@biblabel[1]{#1.}
\makeatother
\bibliographystyle{/home/goyal/Dropbox/Science}

\section*{Acknowledgement}
We thank A. Forbes and F. S. Roux for useful discussions. T.~K. acknowledges the partial support from National Research Foundation of South Africa (Grant specific unique reference number (UID) 86325).
\section*{Author contributions}
S.K.G. and T.K. contributed equally to the content of this article.
\section*{Additional information} The authors do not have competing financial interests.
 

\end{document}